\documentclass[review]{elsarticle}

\usepackage{lineno,hyperref}
\modulolinenumbers[1]

\journal{arXiv}

 \biboptions{numbers,sort&compress}

\usepackage{graphicx}
\usepackage{epstopdf}
\usepackage{amsmath}
\usepackage{amsfonts}
\usepackage[compact]{titlesec}
\usepackage{amsthm}
\usepackage{amssymb}
\usepackage{multirow}
\usepackage{color}
\usepackage{dirtytalk}
\usepackage{mathtools}
\usepackage{geometry}
\usepackage{times}
\usepackage{algorithm}
\usepackage{algorithmicx, algpseudocode}

\usepackage{subcaption}

\usepackage{verbatim}

\theoremstyle{definition}

\newcommand{\bfs}[1]{{\boldsymbol #1}}

\definecolor{Green}{rgb}{0,.5,0}
\definecolor{Blue}{rgb}{0,.1,.85}
\definecolor{Cyan}{rgb}{.2,.6,.7}
\definecolor{Purple}{rgb}{.5,0,1}
\definecolor{deepred}{rgb}{.8,.1,.2}
\newcommand{\rev}[1]{\textcolor{black}{#1}}

\begin{document}

\begin{frontmatter}

\title{Physics-Informed Neural Networks for Discovering Localised Eigenstates in Disordered Media} 

\author[ad]{Liam Harcombe}
\ead{Liam.Harcombe@anu.edu.au}

\author[ad]{Quanling Deng \corref{corr}}
\cortext[corr]{Corresponding author}
\ead{Quanling.Deng@anu.edu.au}

\address[ad]{School of Computing, Australian National University, Canberra, ACT 2601, Australia}

\begin{abstract}
The Schrödinger equation with random potentials is a fundamental model for understanding the behavior of particles in disordered systems. 
Disordered media are characterised by complex potentials that lead to the localisation of wavefunctions, also called Anderson localisation.
These wavefunctions may have similar scales of eigenenergies which poses difficulty in their discovery.
It has been a longstanding challenge due to the high computational cost and complexity of solving the Schrödinger equation.
Recently, machine-learning tools have been adopted to tackle these challenges.  
In this paper, based upon recent advances in machine learning, we present a novel approach for discovering localised eigenstates in disordered media using physics-informed neural networks (PINNs). 
We focus on the spectral approximation of Hamiltonians in one dimension with potentials that are randomly generated according to the Bernoulli, normal, and uniform distributions.
We introduce a novel feature to the loss function that exploits known physical phenomena occurring in these regions to scan across the domain and successfully discover these eigenstates, regardless of the similarity of their eigenenergies.
We present various examples to demonstrate the performance of the proposed approach and compare it with isogeometric analysis. 

\end{abstract}


\end{frontmatter}

\section{Introduction and Background}

Partial differential equations (PDEs) are powerful tools for studying dynamic systems and have applications across various fields. 
Despite their usefulness, these equations are notoriously difficult to solve explicitly, leading to an increased interest in numerical approximations. 
In addition to classical numerical methods, recent research has shown the potential of using neural networks to estimate solutions to differential equations \cite{lagaris1998, lagaris1997artificial, magill2018neural}.
There are several advantages of using neural networks to solve PDEs over traditional numerical methods. 
\rev{
Firstly, numerical errors are usually not accumulated \cite{mattheakis2022}.
Secondly, they are more robust against the ``curse of dimensionality," which refers to the difficulty of solving PDEs in high-dimensional spaces using traditional numerical methods \cite{han2018highDimensional}.
Furthermore, they can be trained to solve a wide range of PDEs, including those with complex boundary conditions or nonlinear operators \cite{flamant2020solving}.
}
Once trained, neural networks can quickly produce solutions for new PDEs that are similar to those encountered during training, potentially without requiring further adjustments or recalibration of parameters \cite{desai2022oneshot}.
Overall, neural networks can provide a powerful and flexible tool for solving PDEs, especially in cases where traditional numerical methods may be insufficient or computationally expensive.

Data-driven supervised networks \cite{finol2018cnn} and data-free unsupervised networks \cite{jin2022pinn} have been shown to produce efficient approximations of differential eigenvalue problems.
Supervised networks aim to discover patterns in labelled datasets to infer a general model that is able to predict examples outside of the given dataset. 
The training process involves calculating approximations for given inputs and computing how much these estimations deviate from the true outputs using a loss function. Then, the networks work to minimise this loss function by adjusting its estimation process. Artificial Neural Networks (ANNs) are comprised of layers of interconnected nodes with weights and biases that calculate the estimations. Convolutional Neural Networks (CNNs) are a newer form of neural networks that are especially known for their ability to identify patterns faster and more reliably than ANNs in solving eigenvalue problems using labelled datasets \cite{finol2018cnn}. They introduce a convolutional layer built up of windows that slide around the input data, searching for patterns in whole regions at a time. They are commonly used to pass information on to ANNs and are optimised similarly. In situations where training data is unavailable or difficult to obtain, unsupervised networks are employed as a solution, operating without a labelled dataset. \rev{In these situations, more attention is drawn to the design of the network's loss function, as there is no labelled data to compare predictions with \cite{chen2022semisupervised}.}


In the case of finding the eigenstates of Hamiltonians, we leverage known physical phenomena to design our loss function. This kind of model is an example of a Physics Informed Neural Network (PINN, c.f., \cite{raissi2019physics,cai2021physics}), one that embeds the knowledge of physical laws governing the system into the learning process to drive the network to an admissible approximation.


Differential eigenvalue problems occur in a wide range of problems in physics and applied mathematics, such as quantum energy problems. 
In early works, Lagaris et al. \cite{lagaris1997artificial} proposed an ANN to discover the solutions to differential eigenvalue problems, tested on various problems in quantum mechanics. 
More recently, Finol et al. \cite{finol2018cnn} presented a supervised ANN to solve eigenvalue problems in mechanics, 
showing that CNNs were outperforming traditional ANNs in contexts with labelled datasets. 
Sirignano and Spiliopoulos \cite{sirignano2018dgm} developed a deep learning network to accurately solve partial differential equations in dimensions as high as 200. 
They also proposed a mesh-free algorithm, which was desirable as meshes become infeasible in such high dimensions. 
Chen et al. \cite{chen2020inverse} adopted the emerging PINNs to solve representative inverse scattering problems in photonic metamaterials and nano-optic technologies. 
Their method was also mesh-free and was tested against numerical simulations based on the finite element method.
These ANN models are excellent at learning time series data.
The work \cite{mano2021lstm} demonstrated its application in wave packet displacements in localised and delocalised quantum states. 
Another work is \cite{kotthoff2021class}, where they adopted CNNs to solve a classification problem. 
The authors constructed a supervised model learning from experimental data to distinguish between the dynamics of an Anderson insulator and a many-body localised phase. 
Yang et al. \cite{yang2021bpinn} proposed a combination of PINNs with Bayesian Neural Networks, which solved both forward and inverse nonlinear problems, obtaining more accurate predictions than PINNs in systems with large noises.


Jin et al. \cite{jin2022pinn} have shown that PINNs can solve single and multiple square well problems, calculating the eigenfunction and eigenvalue simultaneously. Their network employs a scanning mechanism that pushes the network to search for higher eigenvalues as the training process evolves, while the network updates the eigenfunction prediction accordingly. They store found eigenfunctions and exploit the orthogonality of wave functions to search for higher eigenstates. 
Grubišić et al. \cite{Grubisic2021loc} applied PINNs to identify localised eigenstates of operators with random potential. 
They studied the effective potential of the operator, whose local minima correspond to the different eigenstates. 
They first built a PINN to solve these eigenvalue problems in one dimension, 
then generalised to a deeper model that solves higher dimensional problems. 
Effective potentials provide a neat way of locating the localised eigenstates, however, are limited in the number of locations they can identify, 
and have difficulty differentiating between eigenstates of identical eigenvalue. 
These limitations occur in Bernoulli distributed potentials, which we aim to solve with our model.

We build on the design of \cite{jin2022pinn} to approximate these eigenstates for Hamiltonians whose potential is distributed randomly, leading to localisation of the solution to specific regions of the domain. In particular, the core of our work is in finding these eigenstates where the eigenenergies are nearly identical, a task that current models are unable to accomplish. This is most prevalent with potentials distributed according to the Bernoulli distribution but can occur in other distributions.

The rest of the paper is organised as follows. 
In Section \ref{sec:ps}, we state the Schrödinger differential eigenvalue problem and present the isogeometric analysis, followed by a discussion on the challenges of its spectral approximation. 
In section \ref{sec:met}, we propose our novel loss function design for the PINN for solving the Schrödinger equation. 
Section \ref{sec:num} collects and discusses various numerical tests to demonstrate the performance of the proposed method. 
Concluding remarks are presented in section \ref{sec:conc}.

\section{Problem Statement and Isogeometric Analysis} \label{sec:ps}

In this section, we first introduce the modeling problem, followed by a presentation of a classic numerical method, namely isogeometric analysis, used to solve the problem. 
We then provide a general discussion on the challenges associated with the numerical computation of the model.

\subsection{Problem Statement}
We study the time-independent Schrödinger equation: Find the eigenpair $(E, u)$ such that
\begin{equation} \label{eq:pdeSchro} 
\begin{aligned}
\left[-\frac{\hbar^2}{2m} \Delta + V \right] u & = E u \quad  \text{in} \quad \Omega, \\
u & = 0 \quad \text{on} \quad \partial \Omega,
\end{aligned}
\end{equation}
where $\Delta = \nabla^2$ is the Laplacian, $V = V(x) \in L^2(\Omega)$ specifies the potential and is a non-negative function, and $\Omega \subset  \mathbb{R}^d, d=1,2,3,$ is a bounded open domain with Lipschitz boundary $\partial \Omega$. 
\rev{Throughout the paper, we will focus on the case with $d=1$, followed by an example of an extension to $d=2$ in Section \ref{sec:2d}.}
The differential operator is referred to as the Hamiltonian, i.e., $\mathcal{H} = -\frac{\hbar^2}{2m} \Delta + V$. 
Here, $\hbar$ is the reduced Planck constant and $m$ is the mass of the particle. 
\rev{$\frac{\hbar^2}{2m}$ is a constant that we can divide throughout equation \ref{eq:pdeSchro} and absorb into $V$ and $E$. Thus, finding the eigenpairs in equation \ref{eq:pdeSchro} is equivalent to finding those for the following equation, up to a scalar:}
\begin{equation} \label{eq:pde}
\begin{aligned}
- \Delta u + Vu & = E u \quad  \text{in} \quad \Omega, \\
u & = 0 \quad \text{on} \quad \partial \Omega.
\end{aligned}
\end{equation}

This problem is a \rev{Sturm-Liouville eigenvalue problem ($- \Delta + V$ is a diffusion-reaction operator, see, for example, \cite{strang1973analysis,evans2010partial}) which} has a countable infinite set of positive eigenvalues ${E _j} \in {\mathbb{R}^+}$ 
\begin{equation}
0 < {E _1} < {E _2} \le \cdots \le {E _j} \le \cdots
\end{equation}
with an associated set of orthonormal eigenfunctions ${u_j}$ 
\begin{equation}
({u_j},{u_k}) = \int_{\Omega}  {{u_j}(x){u_k}} (x) \ \text{d} \bfs{x} = {\delta _{jk}},
\end{equation}
where $\delta _{jk}$ is the Kronecker delta which is equal to 1 when $j=k$ and 0 otherwise. 
The set of all the eigenvalues is the spectrum of the Hamiltonian. 
We normalise the eigenfunctions in the $L^2$ space; hence, the eigenfunctions are orthonormal under the scalar inner product. 
Let us define two bilinear forms 
\begin{equation} \label{eq:bfs}
a(w,v) = \int_{\Omega}  \nabla w \cdot \nabla v + V w v \ \text{d} \bfs{x} \quad \text{and} \quad b(w,v) = (w, v) = \int_{\Omega} w v \ \text{d} \bfs{x}, \quad \forall w, v \in H^1_0(\Omega),
\end{equation}
where $H^1_0(\Omega)$ is a Sobolev space with functions vanishing at the boundary $\partial \Omega.$ 
Using this notation, the eigenfunctions are also orthogonal with each other with respect to the energy's inner product, i.e.,
\rev{
\begin{equation} \label{eq:oe}
a(u_j, u_k) = {E _j} ({u_j},{u_k})  = {E _j}{\delta _{jk}}.
\end{equation}}
We remark that these orthogonalities are critical in the development of the proposed neural networks in Section \ref{sec:met}.

\subsection{Isogeometric Analysis}

At the continuous level, the weak formulation for the eigenvalue problem \eqref{eq:pde} is: Find all eigenvalues $E \in {\mathbb{R}^+}$ and eigenfunctions $u \in H^1_0(\Omega)$ such that,
\begin{equation} \label{eq:weak}
a(w, u) = E b(w, u), \quad \forall \ w \in H^1_0(\Omega),
\end{equation}
while at the discrete level, the isogeometric analysis (IGA, c.f., \cite{hughes2005isogeometric,cottrell2009isogeometric,deng2018isogeometric,deng2022isogeometric}) for the eigenvalue problem \eqref{eq:pde} is: Find all eigenvalues $E_h \in {\mathbb{R}^+}$ and eigenfunctions $u_h \in W_h$ such that,
\begin{equation} \label{eq:vf}
a(w_h, u_h) = E_h b(w_h, u_h), \quad \forall \ w_h \in W_h(\Omega),
\end{equation}
where $W_h \subset H^1_0(\Omega)$ is the trial and test space spanned by the B-spline or non-uniform rational basis spline (NURBS) basis functions \cite{de1978practical}.

In this paper, for the purpose of comparison with the state of the art, we adopt the recently-developed soft isogeometric analysis (softIGA, c.f., \cite{deng2023softiga,li2023soft}).
SoftIGA has a similar variational formulation as \eqref{eq:vf} which leads to the matrix eigenvalue problem
\begin{equation} \label{eq:mevp}
\mathbf{K} \mathbf{U} = E_h \mathbf{M} \mathbf{U},
\end{equation}
where $\mathbf{K}_{jk} =  a(\phi_j, \phi_k), \mathbf{M}_{jk} = b(\phi_j, \phi_k),$ and $\mathbf{U}$ contains the coefficients of the eigenfunction $u_h$, for its representation in the linear combination of the B-spline basis functions. For simplicity, the matrix $\mathbf{K}$ (although it contains a scaled mass) is referred to as the stiffness matrix while the matrix $\mathbf{M}$ is referred as the mass matrix, and $(E_h, \mathbf{u}_h)$ is the unknown eigenpair. We refer to \cite{deng2023softiga,li2023soft} for details.

\subsection{A Few Challenges on Numerical Computation}

There are a few challenges in the numerical computation, such as using IGA or softIGA, of the Schrödinger equation with random potentials. 
Firstly, there is a non-uniqueness of solutions. The Schrödinger equation with random potentials can have multiple solutions that correspond to the same (or very similar) energy level. 
This non-uniqueness can make it difficult to accurately identify the correct eigenstates, particularly when dealing with complex potential landscapes.
Secondly, it can be a multi-scale problem with large sampling/discretisation errors. 
Random potentials can vary over multiple length scales, which leads to inaccurate numerical simulations and can make it challenging to choose an appropriate discretisation size and mesh.
Thirdly, the Schrödinger equation for many-body problems in multiple dimensions suffers from the curse of dimensionality. 
The number of parameters needed to describe the random potential can be very large. 
As the number of dimensions increases, the computational cost of solving the Schrödinger equation can increase exponentially.

Last but not least, the Schrödinger equation with random potentials is a complex mathematical problem, and solving it numerically can be computationally expensive. 
This is particularly true for large system sizes and high disorder strengths, which require a large number of numerical simulations.
Moreover, solving the Schrödinger equation with a different random potential requires constructing a new matrix eigenvalue problem \eqref{eq:mevp}, which can be a time-consuming process. 
This can become particularly challenging when solving the equation for a large number of random potentials. 
In such cases, one potential solution is to train a neural network that takes the random potential as input and produces the corresponding eigenstates as outputs.
For instance, to solve the Schrödinger equation with $N$ different random potentials, 
one could use a classic method like softIGA and apply it $N$ times, or alternatively, 
train a neural network using data from a subset of cases and then use the well-trained neural network to solve the remaining cases. 
This approach can save a significant amount of computational time, particularly when $N$ is large.
With these challenges in mind, we introduce the following neural-network-based method as an alternative method to solve \rev{the time-independent Schrödinger equations.}

\section{The Physics-Informed Neural Network} \label{sec:met}

Deep neural networks are a type of machine learning algorithm inspired by the structure and function of the human brain. 
They are composed of multiple layers of artificial neurons, each performing a nonlinear transformation of the input data. 
The output of each layer is fed as input to the next layer, allowing the network to learn increasingly complex features and relationships in the data.

Some common types of deep neural networks include convolutional neural networks (CNNs) for image processing and computer vision tasks \cite{lecun1998cnn}, 
long short-term memory models (LSTMs) for sequential data processing \cite{hochreiter1997lstm}, deep belief networks (DBNs) for unsupervised learning \cite{hinton2006dbn}, 
and generative adversarial networks (GANs) for image and video synthesis \cite{goodfellow2014gan}. 
Deep neural networks have demonstrated impressive performance in many applications.
In this paper, we adopt the physics-informed neural networks (PINN) with a novel loss function design to solve the Schrödinger equations.

\subsection{PINN}

\rev{
The Physics-Informed Neural Networks (PINNs) are a type of feed-forward neural networks (FNNs) that incorporate physics laws, typically in the form of partial differential equations (PDEs), into their loss functions \cite{lu2021deepxde,raissi2019physics}. This combination of neural networks and PDEs makes PINNs a powerful tool for solving complex PDEs in various scientific and engineering applications.
The overall process of using PINNs to solve a PDE can be summarized as follows. First, a neural network architecture is constructed. This involves defining the structure of the network, including the number and activation functions of its layers, as well as the number of neurons in each layer.
Next, the loss function is defined. The loss function quantifies the discrepancy between the neural network's predictions and the actual solution to the PDE. In the case of PINNs, the loss function incorporates terms that enforce the PDE itself, as well as any associated initial or boundary conditions.
Once the neural network and loss function are established, the network is trained using an optimization algorithm, such as the Adam optimiser \cite{kingma2014adam}. During the training process, the neural network adjusts its weights and biases to minimise the loss function, gradually improving its ability to approximate the PDE solution.
After training, the accuracy of the neural network's predictions can be evaluated by comparing them against calculations made by current numerical methods, such as isogeometric analysis \cite{cottrell2009isogeometric}. If the predictions are not satisfactory, further refinement can be performed. This may involve modifying the network architecture, adjusting the loss function, or selecting a different optimization algorithm. The refined network is then retrained to improve its performance. Our PINN implementation is summarised in Algorithm \ref{algo:pinn}.
}

Typically, the loss function in PINN is defined as
\begin{equation} \label{eq:loss}
L = L_\text{de} + L_\text{reg},
\end{equation}
where $L_\text{de}$ specifies the loss associated with the PDE and $L_\text{reg}$ specifies the loss associated with the regularisation such as the boundary conditions. 
The loss function term $L_\text{de}$ involves derivatives which are usually evaluated by the automatic differentiation by Tensorflow \cite{abadi2016tensorflow} or by Pytorch \cite{paszke2017automatic}. 
In the following subsection, we present a goal-oriented loss function for the best performance in solving the Schrödinger equations with random potentials. A scale factor is assigned to each regularisation term in $L_\text{reg}$. These scale factors are considered hyperparameters and are tuned to emphasise certain constraints that the network should adhere to in its prediction.

\subsection{Loss Function Design}

To demonstrate the main idea, we simplify our focus to the one-dimensional case.
We assume that the potential $V(x)$ is randomly distributed over $m$ regions over the domain $[0, 1]$, and each region has a random value. 
Each set of random values then characterises the individual differential eigenvalue problem to be solved for Anderson localised states. 
We partition the interval $[0,1]$ into $n+1$ equally spaced points $x_j = jh, j=0,1,\cdots,n$ where $h=1/n$ is the grid/mesh size. 
We denote by $\bfs{u}_h = (u_h(x_0), u_h(x_1), \cdots, u_h(x_n) )^T$ as a vector of the approximate values of the true solution $u(x)$ at each of these coordinates $x_j$.
Similarly, we denote by $\bfs{V} = (V(x_0), V(x_1), \cdots, V(x_n) )^T$ as a vector of the approximate values of the potential $V(x)$.
In our PINN model, we choose $n$ such that each region has exactly 5 nodes each for accuracy as well as for training efficiency. 
To approximate the second derivative vector $u''(x)$, we use the center finite difference method with an accuracy order of six, i.e., the error is of $\mathcal{O}(h^6)$.

\begin{figure}[ht]
 	\centering
 	\includegraphics[scale=0.4]{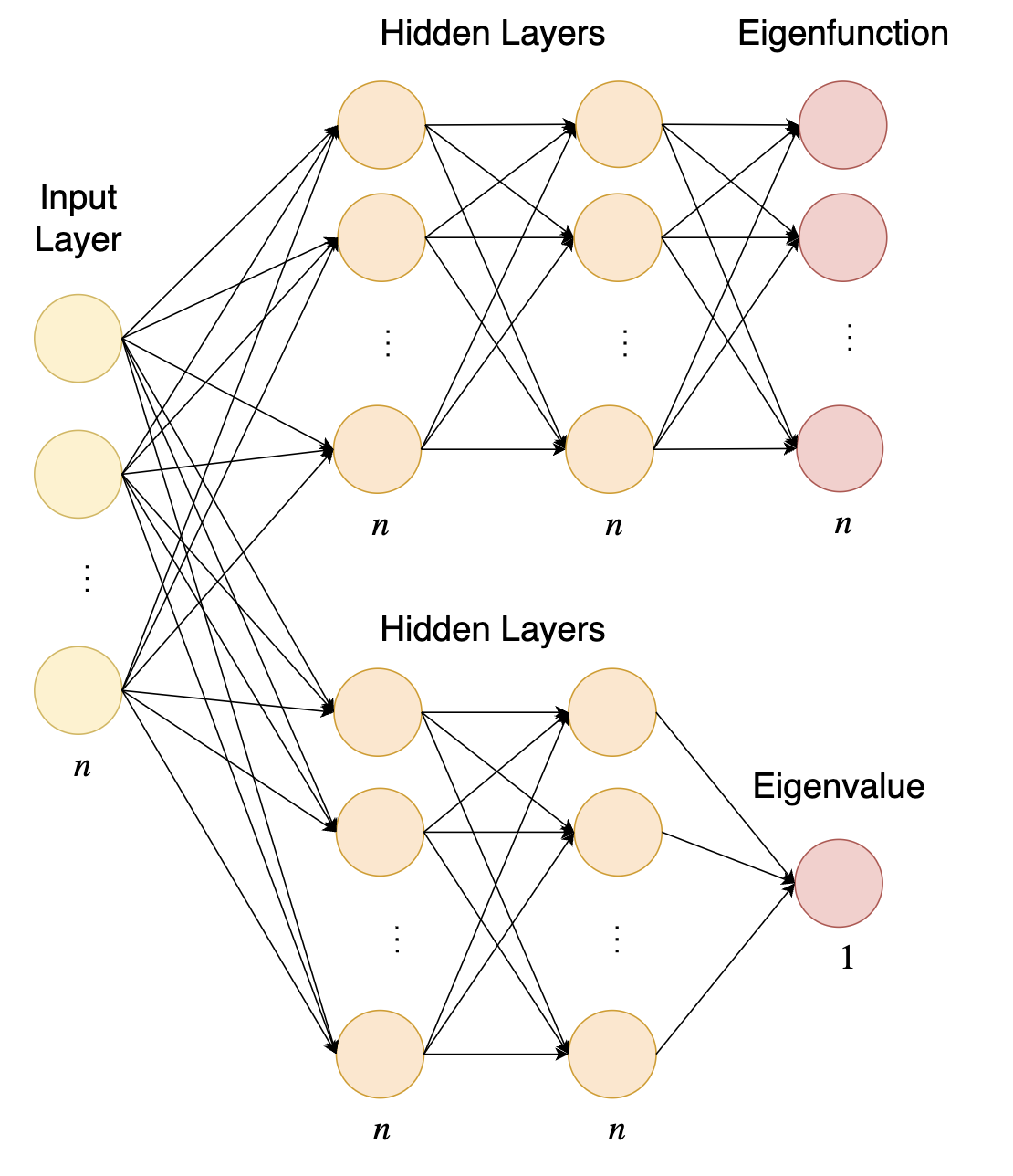}
 	\captionof{figure}{Neural network structure for the prediction of eigenstates of Hamiltonians.}
 	\label{fig:nn}
 \end{figure}

\rev{
We construct a PINN that splits into two separate networks: 
one that calculates the eigenvalue $E$, and the other one produces the eigenfunction $u(x)$ at certain nodes. 
These two networks are trained and optimised simultaneously, with their outputs used in the same loss function described below. We consider the combination of these two networks as one network and apply it to discover the eigenstates.
Figure \ref{fig:nn} draws this model structure, highlighting the split into separate networks. 
The network section finding the eigenvalue has two hidden layers with $n$ nodes, while the network for finding the eigenfunction has three hidden layers with $n$ nodes. This allows the network to scale with the size of the mesh.
Each layer in both networks uses the ReLU activation function $f(x)=$ max($x, 0$), and the loss minimisation is done using the Adam optimiser with a learning rate of $5\times 10^{-4}$.}

The driving force of this network towards a correct solution is the design of its loss function, which is built of multiple terms relating to different aspects of a correct solution. The neural network's optimisation function works to minimise the loss function, so we choose terms that are positive errors for incorrect predictions of $E$ and $u(x)$, and are zero for correct predictions. Thus, the minimisation of these terms leads the network to produce correct predictions for $E$ and $u(x)$. \rev{During training, we choose to input an initial vector of ones (of length $n$), which the network uses as a basis to build its predictions of $E$ and $u(x)$ from.}

The central term of our loss function is the term:
\begin{equation}
L_{\text{de}} = \sqrt{\frac{1}{n+1}\sum_{i=0}^n\left(E_h u_h(x_i)+u_h''(x_j)-V(x_j)u_h(x_j)\right)^2}
\end{equation}
which is always non-negative and is zero for solutions $E$ and $u(x)$ that \rev{satisfy equation \ref{eq:pde}. }
To encourage the network to satisfy the boundary conditions $u(0)=u(1)=0$, we create the term:
\begin{equation}
L_{\text{bound}} = u_h^2(x_0) + u_h^2(x_n)
\end{equation}
which the minimisation of the loss function leads to $u_h(x_0)=u_h(x_n)=0$. 
The major issue with using just these terms is that the trivial solution $u(x)=0$ satisfies these terms (reducing them to zero). Thus, we need the following term to penalise the approximate solution:

\begin{equation} \label{eq:lnorm}
L_{\text{norm}} = \left(\int_0^1u^2_h(x) \text{d} \bfs{x} - 1 \right)^2.
\end{equation}
The minimisation of the loss function leads to approximate solutions of $L^2$ norm 1.
Integration is calculated using the mid-point rule and this calculation is majorly local.
In the PINN model developed in \cite{jin2022pinn}, this condition was imposed using the terms like $1/u^2(x)$ and $1/E^2$.
We point out that since the eigenstates are localised, an eigenstate $u(x)=0$ for most of the domain, leads to very large and even infinite values due to the design of the term $1/u^2(x)$. Consequently, this leads to difficulties in minimising the loss function and the overall neural network training.

These terms are sufficient to push the network to generate an admissible eigenstate. 
However, the model often produces the same eigenstates, preventing the discovery of higher states. 
To overcome this issue, once an eigenstate $(E_h, u_h(x))$ is discovered \rev{(using a patience condition described below)}, 
we add the eigenstate $u_h(x)$ to a list of eigenstates $S$. 
\rev{Then, we stop the training, recompile the network and begin training again, where we take advantage of the known physical fact that the eigenfunctions of Hamiltonians are orthogonal and incorporate the following term into the loss function:}

\begin{equation}
L_{\text{orth}} = \sum_{\hat{u}_h (x)\in S} \left(\int_0^1 \hat{u}_h(x) u_h(x)dx\right)^2,
\end{equation}
where $u_h(x)$ is the network's current state to be produced. 
\rev{We iteratively repeat this process of adding eigenstates to $S$ and recompiling the network. This iteration is automated and can be done an arbitrary number of times. Sometimes the network will skip an eigenstate and converge to a higher one, however, with enough iterations the network tends to find these states eventually.}
To decide whether a prediction is admissible, we determine whether $L_{\text{de}}$ is below a certain threshold, 
as well as a patience condition as in \cite{jin2022pinn} that checks if the absolute value of the change in $L_{\text{de}}$ and that of the eigenvalue is below another threshold. 
This allows the network to converge to solutions with $L_{\text{de}}$ much lower than the chosen threshold, given that it is converging fast enough.


The loss function we have built so far performs adequately on problems with enough spread in the eigenvalues between different states. However, in situations where the eigenstates have nearly identical eigenvalues, the network is unable to distinguish between the states, 
converging to an undesirable linear combination of the eigenfunctions. 
\rev{This occurs because eigenspaces form vector spaces, so a linear combination of eigenfunctions of the same eigenvalue $E$ will be another eigenfunction of eigenvalue $E$, giving another eigenstate that would satisfy $L_\text{de}$ and all the regularisation terms defined.}
When studying Hamiltonians with potentials generated randomly, there is a disorder for Anderson localisation to occur. 
We leverage this fact in our loss function with the following procedure.

\begin{figure}[ht]
     \centering
     \includegraphics[width=0.45\textwidth]{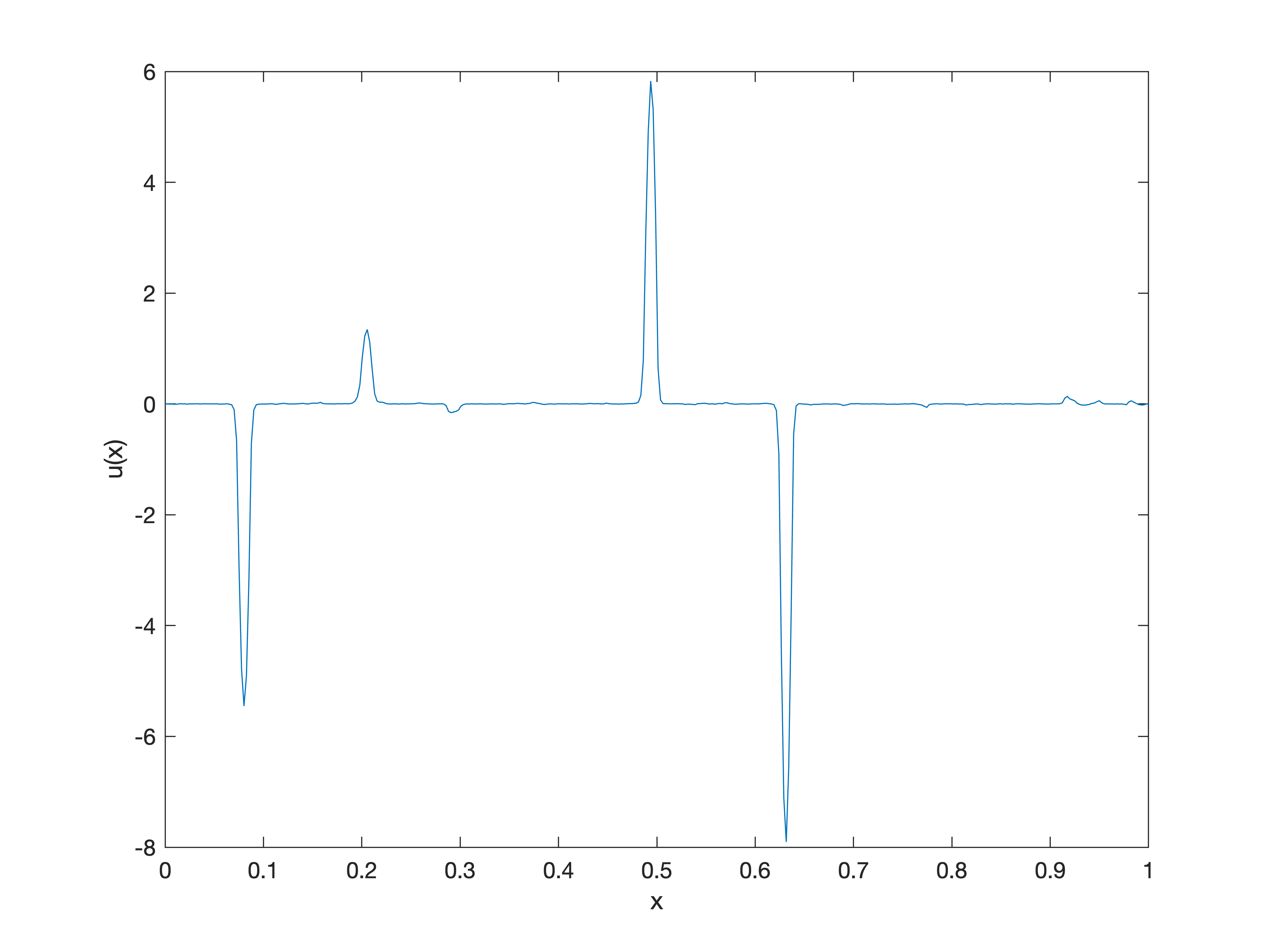}
     \includegraphics[width=0.44\textwidth]{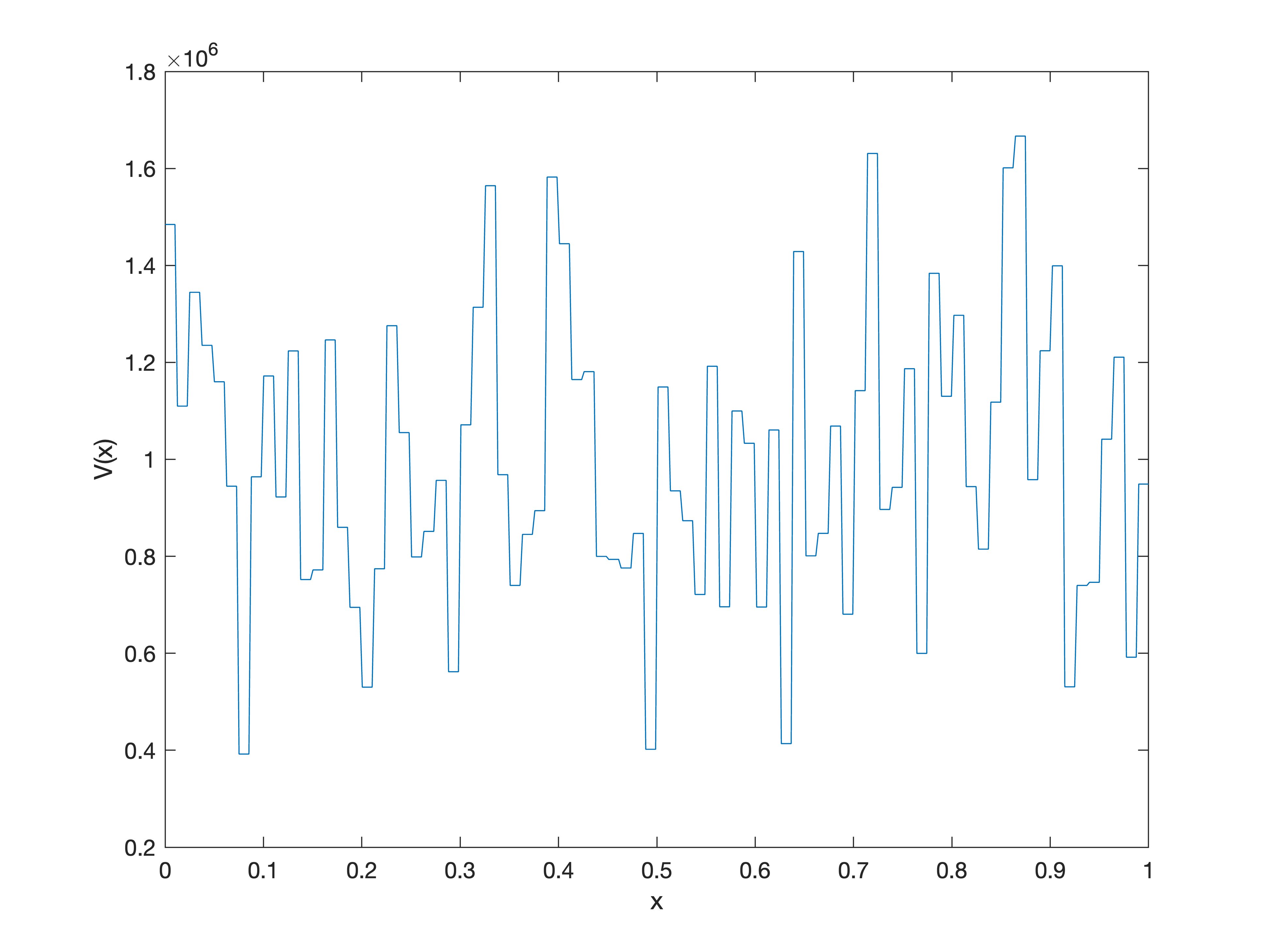}
     \caption{Left: An example of a linear combination of eigenstates where their eigenenergies are approximately equal. 
     Right: The corresponding potential $V(x)$ with $m=80$ uniform elements, where $V(x)$ was randomly generated in each element according to the normal distribution with mean 1 and standard deviation 0.3, scaled by $10^6$. }
     \label{fig:u}
 \end{figure}

We let the model run with the above loss function for a certain number of epochs $q$, which is a hyperparameter. 
The goal is to set this parameter so that at epoch $q$, the network has converged to a linear combination of the eigenfunctions, for example in Figure \ref{fig:u}.

Since we know the eigenstates of the Hamiltonian are localised,
 each spike in \ref{fig:u} represents a separate localised state. 
 We thus scan through the eigenfunction and generate a list $K$ of the regions of the domain where the norm of the eigenfunction is greater than some set hyperparameter, splitting apart the regions where the function returns to zero to separate the localised spikes. 
 Then, in our iterative process of discovering eigenvalues, at each step, we choose one of the intervals in $K$ (say, $[x_a, x_b]$) and add the following term to the loss function:
\begin{equation}
L_{\text{loc}} = \left( \int_0^1 u^2_h(x) \text{d} \bfs{x} - \int_{x_a}^{x_b} u^2_h(x) \text{d} \bfs{x}\right)
\end{equation}
which encourages the network to set the eigenfunction to zero outside of the localisation interval $[x_a, x_b]$. Equivalently, we could have added terms like $L_{\text{end}}$ for each nodal point outside of $[x_a, x_b]$.
We remark that this term is non-negative. 
\rev{With all these loss terms in mind, we have
\begin{equation}
L_\text{reg} = \alpha_\text{bound}L_{\text{bound}}  + \alpha_\text{norm}L_{\text{norm}}  + \alpha_\text{orth}L_{\text{orth}} + \alpha_\text{loc}L_{\text{loc}},
\end{equation}
which is added to $L_{\text{de}}$ to give the overall loss function $L$ in \eqref{eq:loss}.
The $\alpha$ terms in $L_\text{reg}$ scale each term in the loss function, providing emphasis to certain physical constraints. Through hyperparameter tuning, we found that the following values provided adequate predictions for the number of nodes we used (order of magnitudes $10^2$ to $10^3$)
\begin{equation}
	\alpha_\text{bound} = n^2, \quad \alpha_\text{norm} = n^3, \quad
	\alpha_\text{orth} = n^3, \quad
	\alpha_\text{loc} = n^4.
\end{equation}
The idea of having a large weight on the regularisation terms of the loss function is to ensure the network prioritises minimising these terms before working to a solution satisfying the differential equation (i.e. before minimising $L_\text{de}$). This also penalises the network heavily for straying its solution from the regularisations in $L_\text{reg}$, working to project the parameter search space to a subspace adhering to the regularisations. The method described above is presented in Algorithm \ref{algo:pinn}. The purpose of this algorithm is to discover the eigenstates for a given potential.}
Throughout the paper, we use this loss function in our proposed PINN with the structure shown in Figure \ref{fig:nn}.

\begin{algorithm}
	\caption{Eigenstate Discovery Using PINN}
	\begin{algorithmic}[1]
		\State Let $k$ be the epoch where $L_\text{loc}$ is added to the loss function
		
		\State Let $j = 1$

		\State \textbf{while} $j \leq $ num\_eigenstates \textbf{do}
		
		\State \hspace{\algorithmicindent} Initiate the model
		\State \hspace{\algorithmicindent} \textbf{while} training \textbf{do}
		
		\State \hspace{\algorithmicindent} \hspace{\algorithmicindent} Calculate prediction of $E$ and $u(x)$
		
		\State \hspace{\algorithmicindent} \hspace{\algorithmicindent} Compute $L_\text{de}$, $L_\text{bound}$, $L_\text{norm}$
		
		\State \hspace{\algorithmicindent} \hspace{\algorithmicindent} Compute $L_\text{orth}$ using all stored eigenfunctions
		
		\State \hspace{\algorithmicindent} \hspace{\algorithmicindent} \textbf{if} current epoch $= k$ \textbf{then}
		
		\State \hspace{\algorithmicindent} \hspace{\algorithmicindent} \hspace{\algorithmicindent} Scan predicted eigenfunction for localisation sections
		
		\State \hspace{\algorithmicindent} \hspace{\algorithmicindent} \hspace{\algorithmicindent} Choose the localisation section $S$ with greatest norm
		
		\State \hspace{\algorithmicindent} \hspace{\algorithmicindent} \hspace{\algorithmicindent} Reset model weights 
		
		\State \hspace{\algorithmicindent} \hspace{\algorithmicindent} \textbf{else if} current epoch $> k$ \textbf{then}
		
		\State \hspace{\algorithmicindent} \hspace{\algorithmicindent} \hspace{\algorithmicindent} Compute $L_\text{loc}$ across $S$
		
		\State \hspace{\algorithmicindent} \hspace{\algorithmicindent} \textbf{end if}
		
		\State \hspace{\algorithmicindent} \hspace{\algorithmicindent} Backpropagate and step
		
		\State \hspace{\algorithmicindent} \hspace{\algorithmicindent} \textbf{if} patience condition \textbf{and} $L_\text{de} < $ threshold \textbf{then}
		
		\State \hspace{\algorithmicindent} \hspace{\algorithmicindent} \hspace{\algorithmicindent} Store prediction of $E$ and $u(x)$ for future $L_\text{orth}$ calculation
		
		\State \hspace{\algorithmicindent} \hspace{\algorithmicindent} \hspace{\algorithmicindent} $j = j+1$
		
		\State \hspace{\algorithmicindent} \hspace{\algorithmicindent} \hspace{\algorithmicindent} \textbf{break while} 
		
		\State \hspace{\algorithmicindent} \hspace{\algorithmicindent} \textbf{end if}
		
		\State \hspace{\algorithmicindent} \textbf{end while}
		
		\State \textbf{end while}

	\end{algorithmic}
	\label{algo:pinn}
\end{algorithm}

\section{Numerical Experiments} \label{sec:num}
We test our PINN model on various Hamiltonians with randomly distributed potential 
and demonstrate the network's robustness against potential with different distributions. 
In particular, we consider potentials with distributions that cause the Hamiltonian to admit eigenstates with almost identical eigenenergies. 
This is most prevalent in potentials distributed according to the Binomial distribution. 
\rev{We test the network for different numbers of randomly generated regions ($m$), and choose the number of nodes in the mesh ($n$) such that each region contains 5 nodes (that is, we choose $n=5m$).}

\begin{figure}[ht]
     \centering
     \includegraphics[width=.9\linewidth]{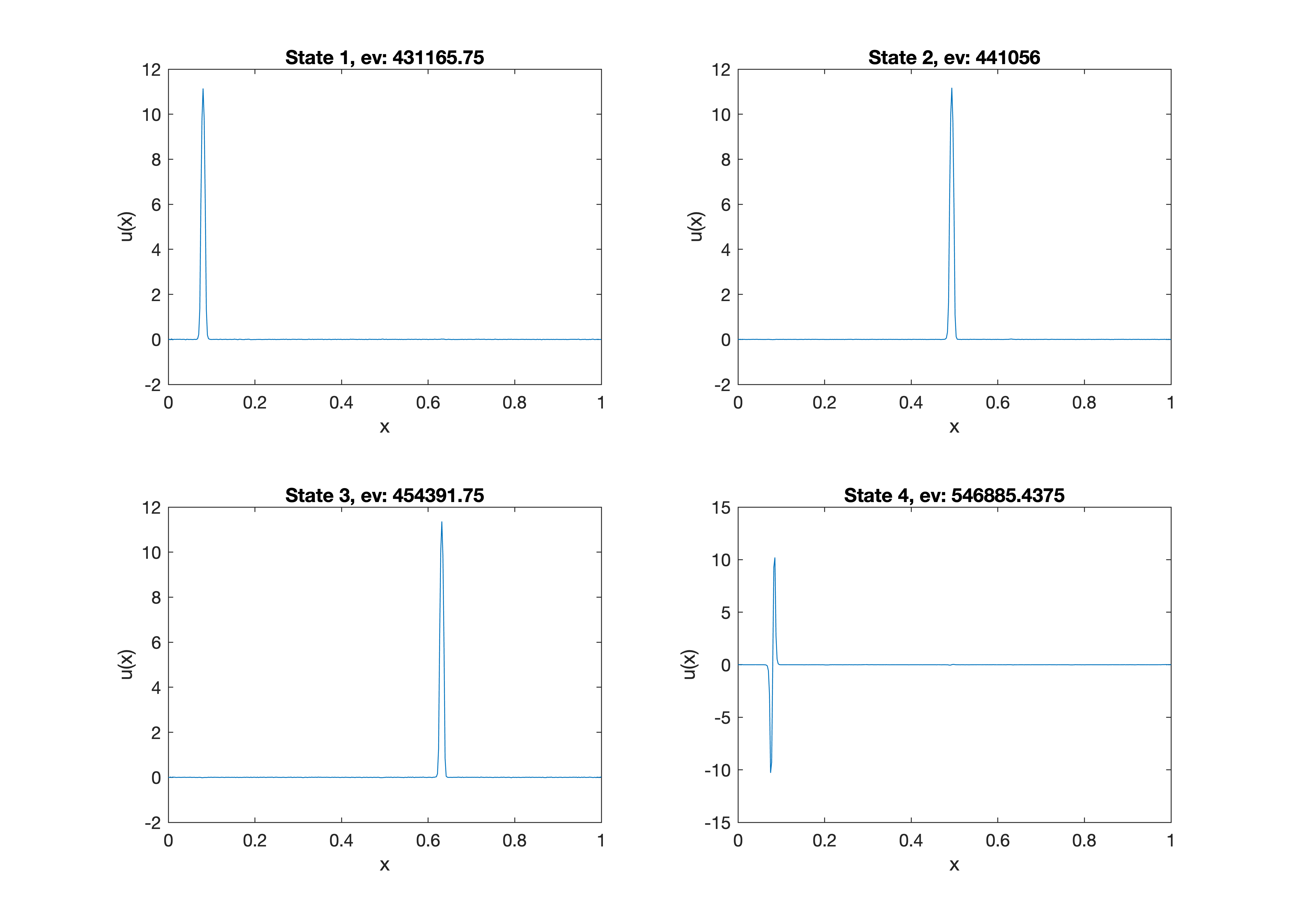}
     \caption{The first four PINN approximated eigenstates with the potential defined in the right plot of Figure \ref{fig:u}. Here, ev refers to the eigenvalue associated with the plotted eigenfunction.}
     \label{fig:normal_400n}
 \end{figure}

The localised states usually occur in the region where the potential takes its smallest value. 
When there are multiple disjoint regions at (or very close to) this minimum value, each of these regions can admit a localised state, 
where they all can have similar eigenenergy. 
Figure \ref{fig:normal_400n} shows the first four PINN approximated eigenstates where the potential is shown in the right plot of Figure \ref{fig:u}.
The network can distinguish between the first three eigenstates, regardless of how similar their eigenvalues are. 

We observe that if $u(x)$ satisfies equation \ref{eq:pde}, then $-u(x)$ also does. $-u(x)$ also has the same $L^2$ norm as $u(x)$, so it satisfies all other terms in our loss function.
\rev{This is relevant as the network will choose to converge to either $u(x)$ or $-u(x)$. For consistency, we plot all eigenfunctions in their positive form.}

 \begin{figure}[ht]
     \centering
     \includegraphics[width=.6\linewidth]{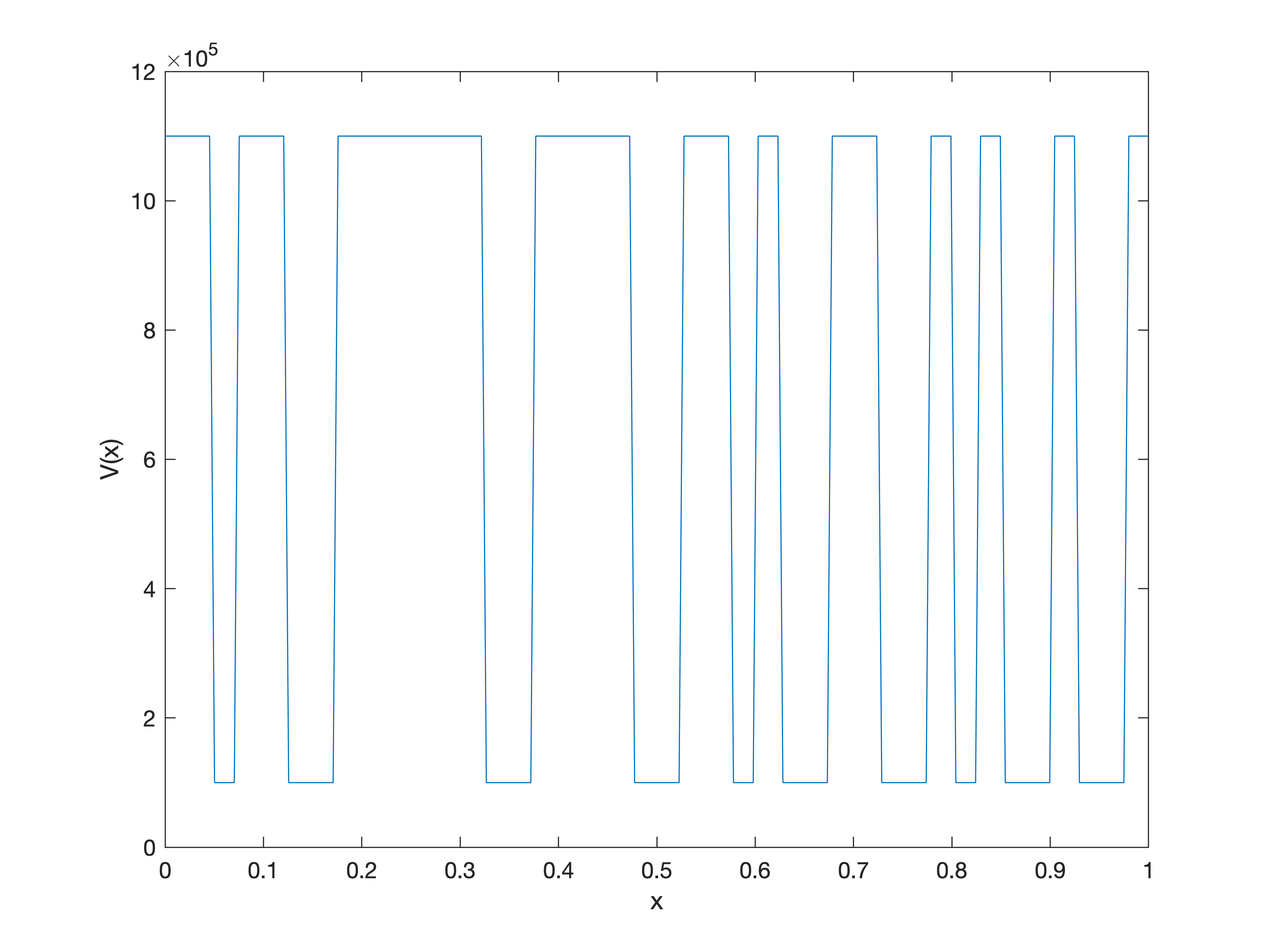}
     \caption{The potential $V(x)$ with $m=40$ uniform elements, where $V(x)$ was randomly generated in each element according to the Bernoulli distribution with probability 0.5, scaled by $10^6$ and shifted up by $10^5$. }
     \label{fig:vb200n}
 \end{figure}

 \begin{figure}[ht]
     \centering
     \includegraphics[width=.9\linewidth]{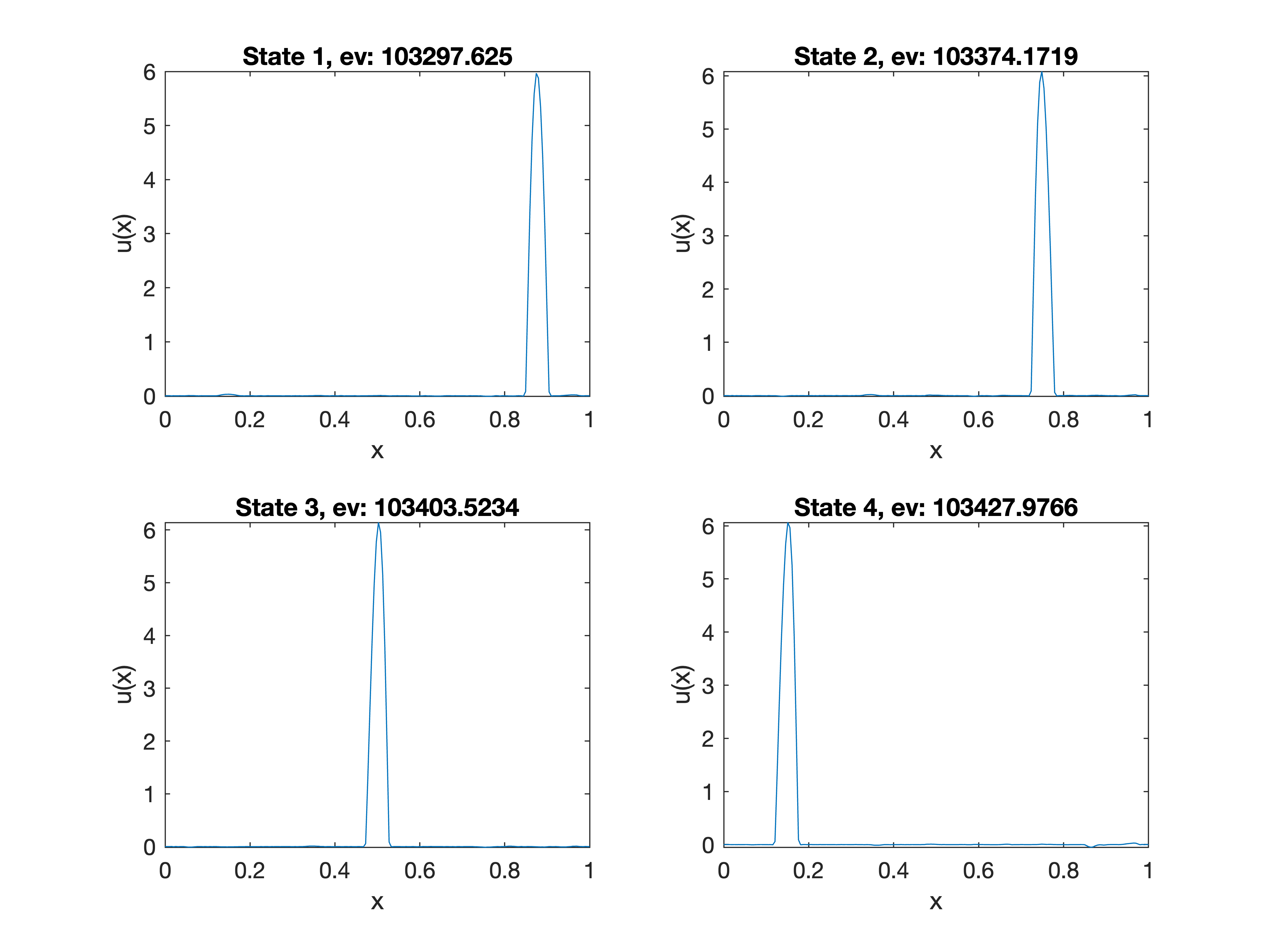}
     \caption{The first four PINN approximated eigenstates with the potential defined in Figure \ref{fig:vb200n}. Here, ev refers to the eigenvalue associated with the plotted eigenfunction.}
     \label{fig:vbu200n}
 \end{figure}

 \begin{figure}[ht]
	\centering
	\includegraphics[width=.6\textwidth]{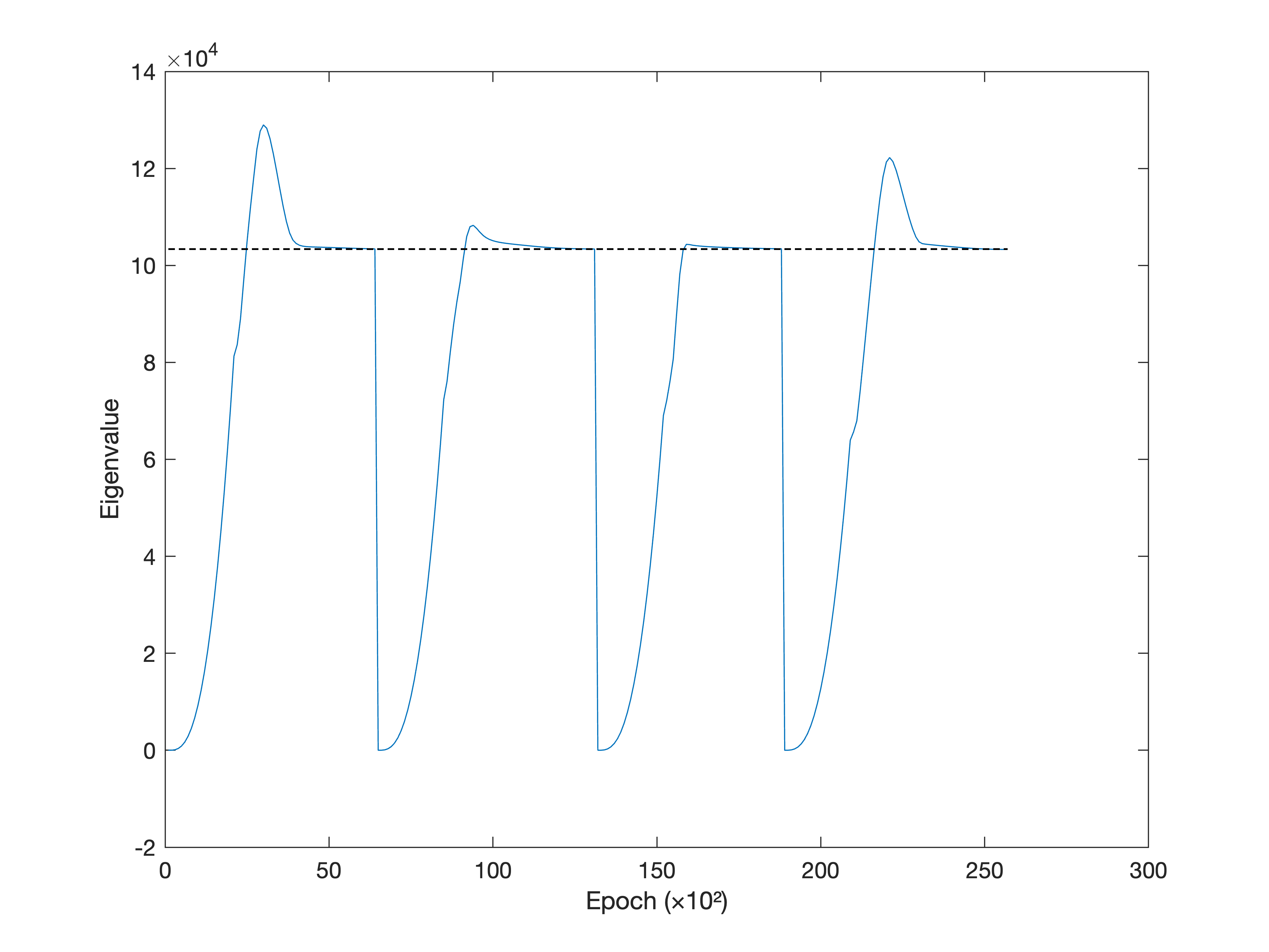}
	\caption{The eigenvalue prediction with respect to the epoch (training process) using the potential in Figure \ref{fig:vb200n}. The dotted lines represent the true eigenvalues.}
	\label{fig:bernoulli_200n_evs}
\end{figure}

Figure \ref{fig:vb200n} shows a potential with a Bernoulli distribution, 
while Figure \ref{fig:vbu200n} shows the first four eigenstates, the first three of which have almost identical eigenvalues. 
We expect these eigenstates to localise in the \say{troughs} of the potential, which is what we observe in the eigenfunction plots. 
The states that localise in troughs of similar length tend to have almost the same eigenvalue, 
which our model captures.

Figure \ref{fig:bernoulli_200n_evs} shows the model's prediction of the eigenvalue as the training process evolves. 
After 2000 epochs, the model adds the $L_{\text{loc}}$ term.
The model produces converging eigenvalues until around epoch 9000. 
At this point, the model's prediction has met the convergence criteria and saves the eigenstate to use in $L_\text{orth}$ 
and it resets its weights to search for the next eigenstate. 
In our implementation, we include a function that generates a video to display the network's eigenfunction prediction as the training process evolves, which is available at our GitHub\footnote{https://github.com/liamharcombe4/eigen-network} page.

Table \ref{table:results} displays the PINN eigenvalue errors where we used the reference eigenvalues generated by the softIGA method with a very fine mesh (2000 nodes). 
We also test the model's robustness against the number of nodes. 
We note that the model achieves slightly better results for the potential distribution according to the normal distribution compared to the uniform distribution. 
However, the results for the Bernoulli distributed potentials are much better, all with relative errors below 1\%.
This demonstrates the success of our approach to the problem of eigenstates with similar eigenvalue.
Table \ref{table:results_bernoulli} displays the eigenvalue prediction results for a single potential generated by the Bernoulli distribution with a different number of mesh elements. 
As the number of elements increases, there is a clear improvement in the accuracy of the PINN-predicted eigenvalues. 
The convergence rate is of order approximately 1 but the theoretical analysis of this convergence is subject to future study.

\begin{table}[h!]
	\centering
	\begin{tabular}{||c c c | c||} 
		\hline
		Potential Distribution & Nodes ($n+1$) & Regions ($m$) & Eigenvalue Error ($\%$)\\ [0.5ex] 
		\hline\hline
            Normal & 400 & 80 & 0.863 \\
            Normal & 200 & 40 & 1.16 \\
            Normal & 100 & 20 & 0.123 \\
            \hline
            Bernoulli & 200 & 40 & 0.351 \\ 
		Uniform & 200 & 40 & 1.92 \\
		\hline
	\end{tabular}
	\caption{The relative eigenvalue errors of the PINN-approximated first eigenenergy of Hamiltonians with potentials generated according to various distributions.}
	\label{table:results}
\end{table}

\begin{table}[h!]
	\centering
	\begin{tabular}{||c | c||} 
		\hline
		Nodes ($n+1$) & Eigenvalue Error ($\%$)\\ [0.5ex] 
		\hline\hline
            100 & 0.783 \\
            200 & 0.343 \\
            300 & 0.0512 \\
		\hline
	\end{tabular}
	\caption{The relative eigenvalue errors of the PINN-approximated first eigenenergy of Hamiltonian with a fixed Bernoulli distributed potential over $m=20$ regions.}
	\label{table:results_bernoulli}
\end{table}

\subsection{\rev{Loss Terms Analysis}}

\begin{figure}[ht]
	\centering
	\includegraphics[width=.6\linewidth]{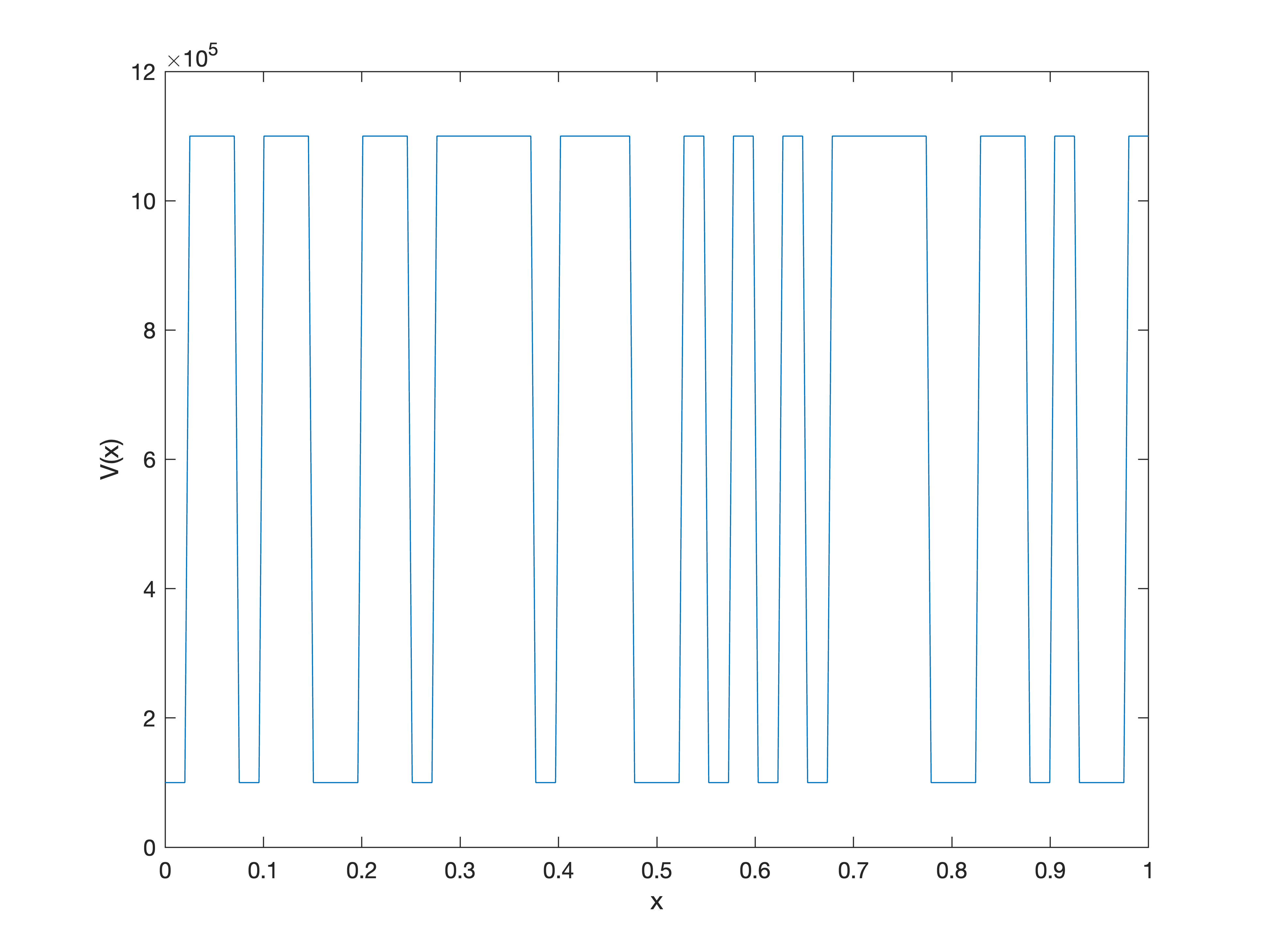}
	\caption{The potential $V(x)$ with $m=40$ uniform elements, where $V(x)$ was randomly generated in each element according to the Bernoulli distribution with probability 0.5, scaled by $10^6$ and shifted up by $10^5$.}
	\label{fig:vb200n2}
\end{figure}

\begin{figure}[ht]
	\centering
	\includegraphics[width=.9\linewidth]{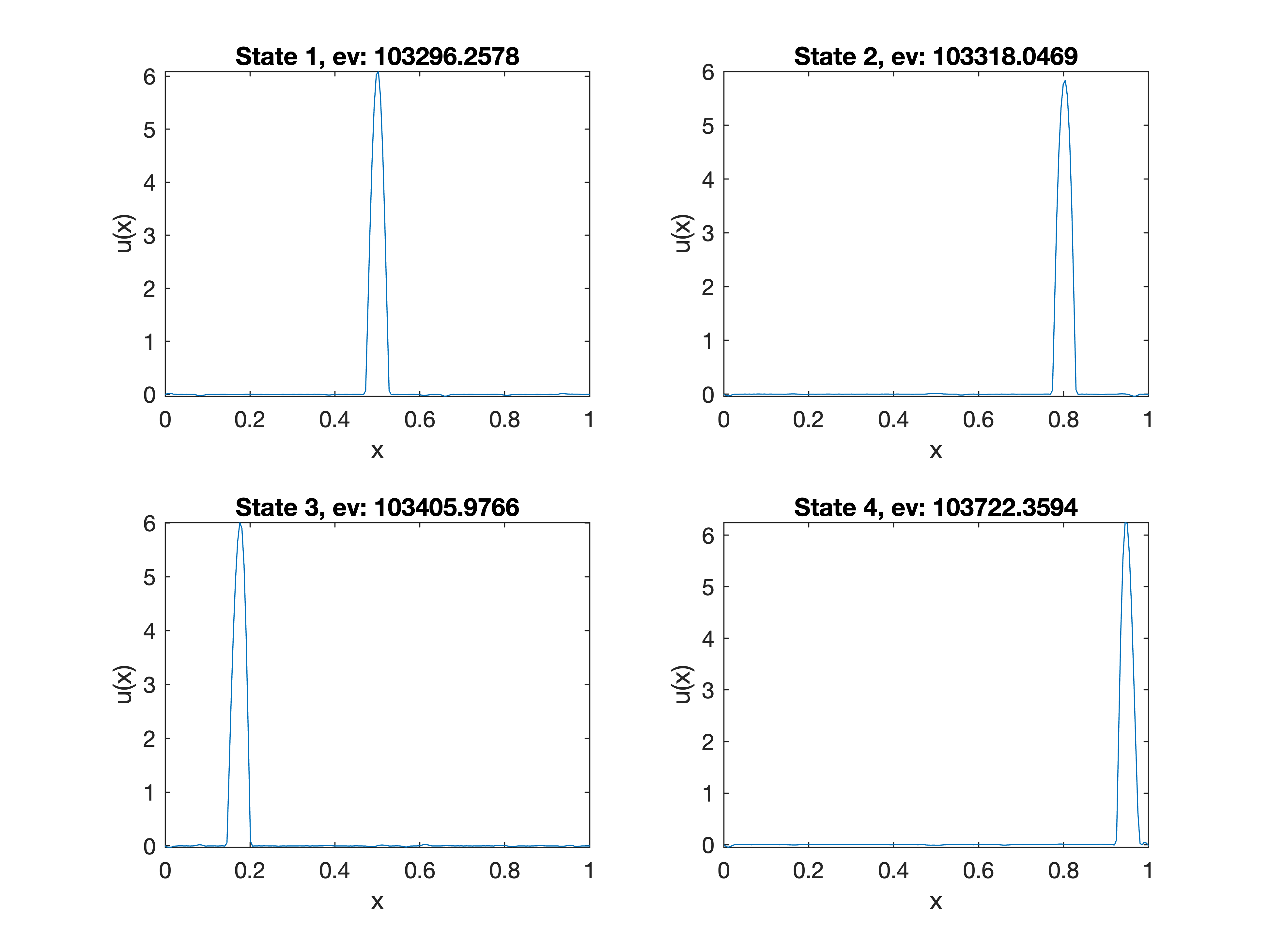}
	\caption{The first four PINN approximated eigenstates with the potential defined in Figure \ref{fig:vb200n2}.}
	\label{fig:vbu200n2}
\end{figure}

\begin{figure}[ht]
	\centering
	\includegraphics[width=\textwidth]{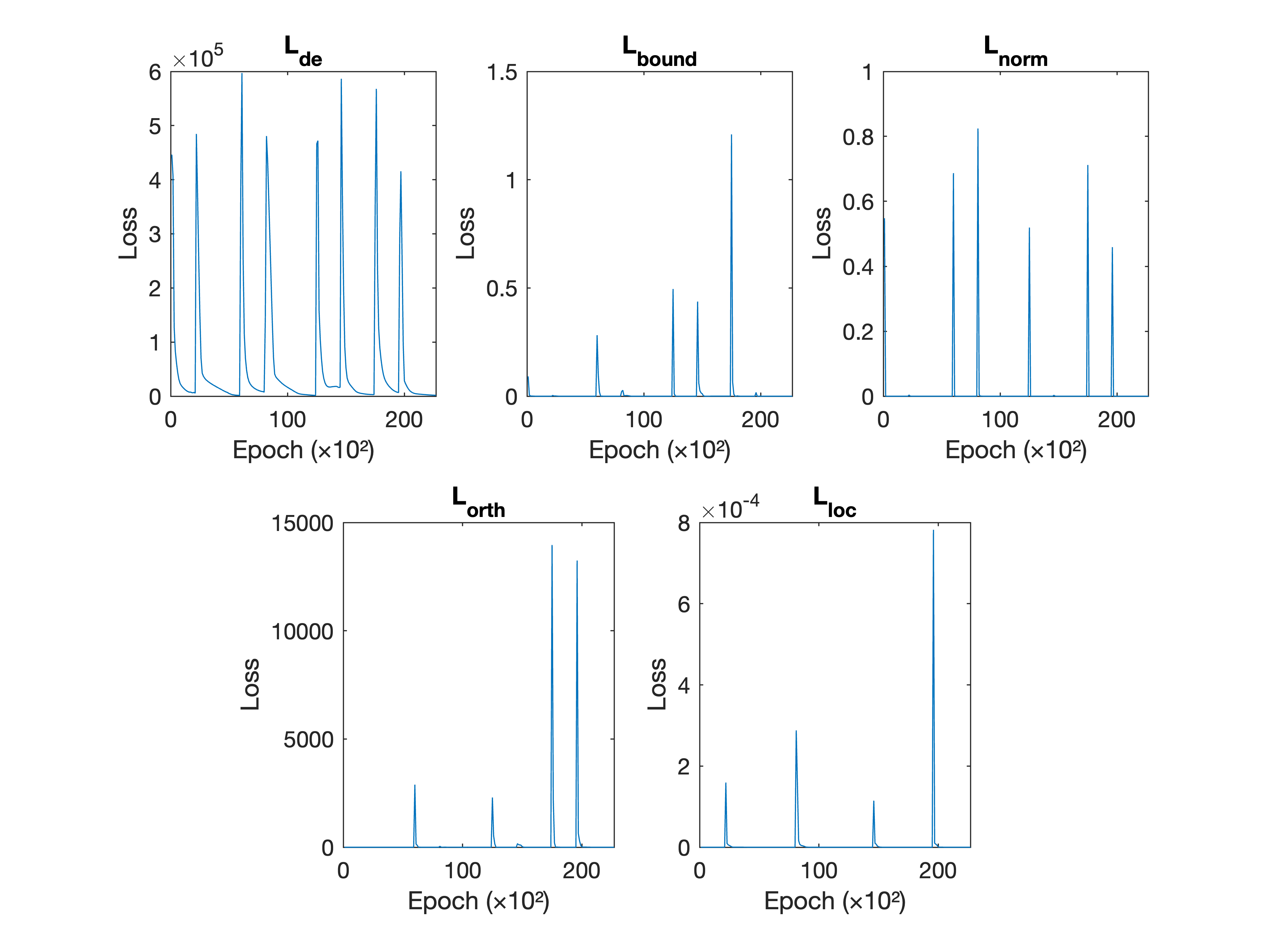}
	\caption{The values of the individual terms in the loss function as the training process evolves.}
	\label{fig:losses}
\end{figure}

\rev{
Another potential based on the Bernoulli distribution is plotted in Figure \ref{fig:vb200n2}, with the network's eigenstate predictions in Figure \ref{fig:vbu200n2}. During the training process of the network for these approximations, the value of each term of the loss function is saved every 100 epochs. Figure \ref{fig:losses} displays the evolution of these terms throughout the training process. 
}

\rev{
The spikes in Figure \ref{fig:losses} represent places in the training process where either the $L_\text{loc}$ term is added to the loss function or the network's parameters are reset after successfully finding an eigenstate and preparing to search for another. We highlight that the regularisation terms are minimised almost instantly, shown by how sharp the spikes are in their plots over training time. This demonstrates that our scaling of the regularisation terms is successfully restricting the network to a search space of solutions that satisfy these regularisation terms, before working to minimise $L_\text{de}$.
}

\subsection{\rev{Comparison with SoftIGA}}

\begin{figure}[ht]
	\centering
	\includegraphics[width=.6\linewidth]{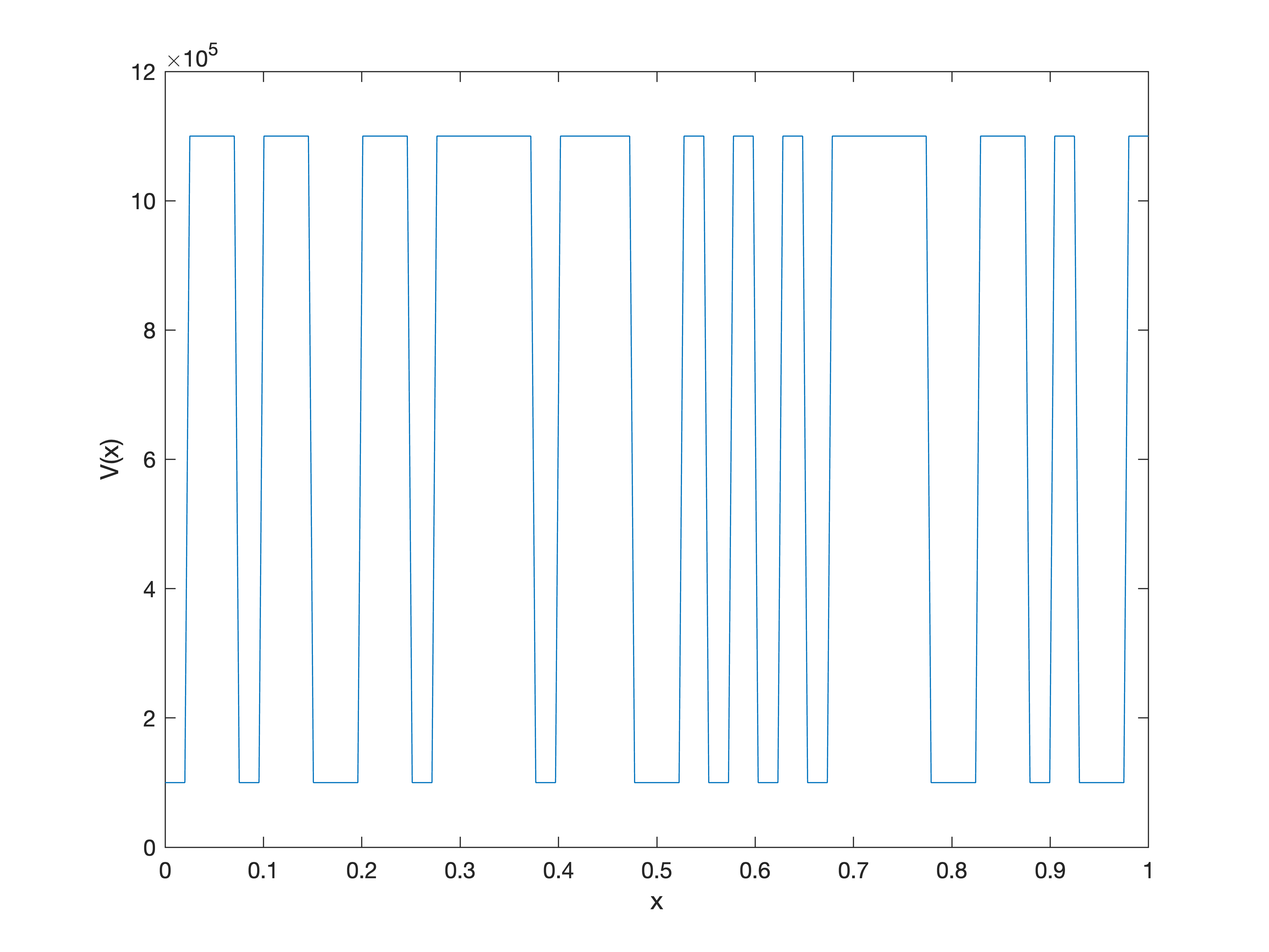}
	\caption{The potential $V(x)$ with $m=40$ uniform elements using $n=200$ nodes, where $V(x)$ was randomly generated in each element according to the Bernoulli distribution with probability 0.5, scaled by $10^6$ and shifted up by $10^5$.}
	\label{fig:iga_pinn_V}
\end{figure}

\begin{figure}[ht]
	\centering
	\includegraphics[width=.9\linewidth]{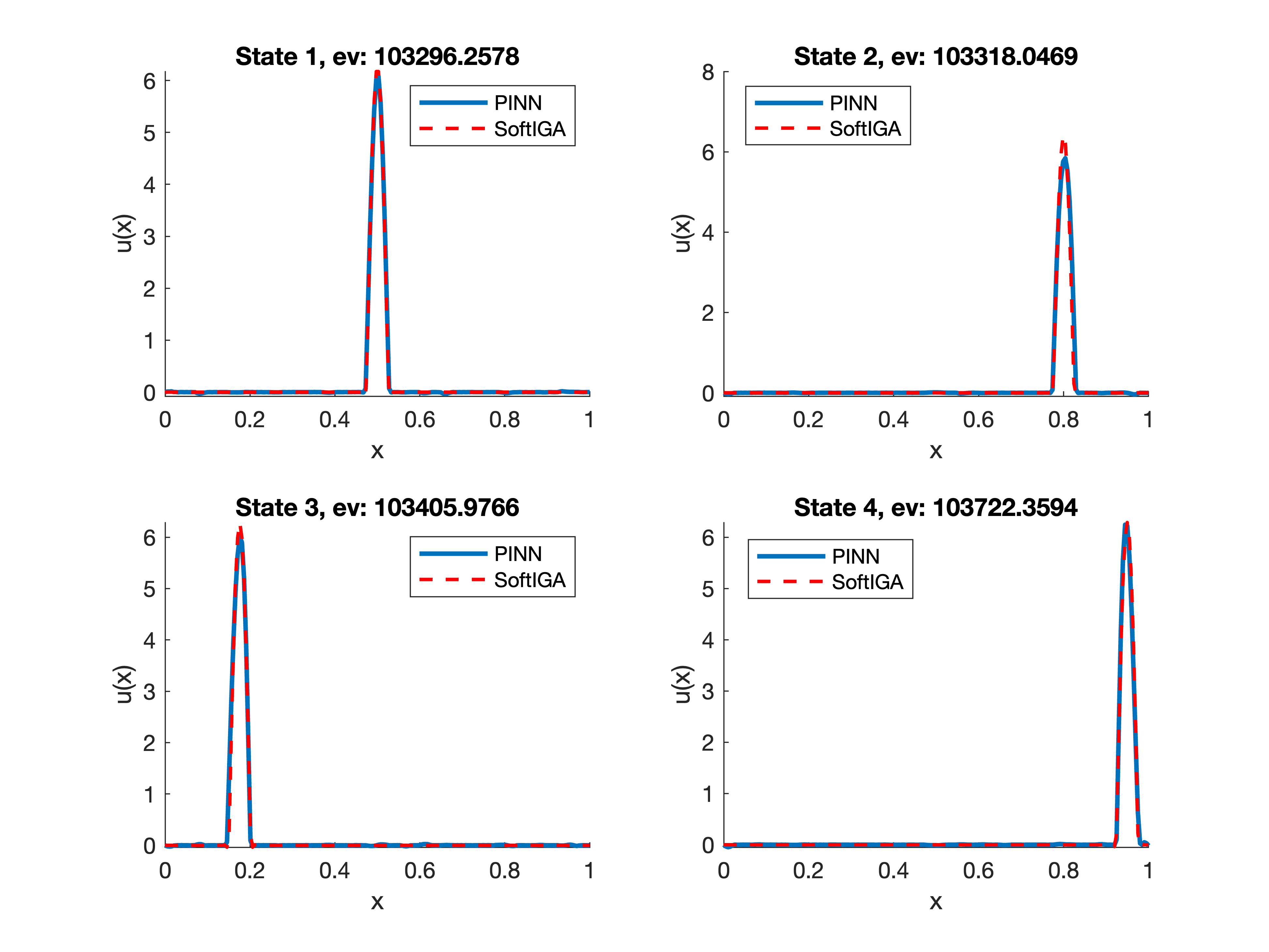}
	\caption{The first four PINN predicted eigenstates compared to the SoftIGA calculated states with the potential defined in Figure \ref{fig:iga_pinn_V}. Here, ev refers to the eigenvalue calculated by the PINN for each state, with the SoftIGA calculated eigenvalue of 103679.96 for each of these four states.}
	\label{fig:iga_pinn_evecs1to4}
\end{figure}

\begin{figure}[ht]
	\centering
	\includegraphics[width=.9\linewidth]{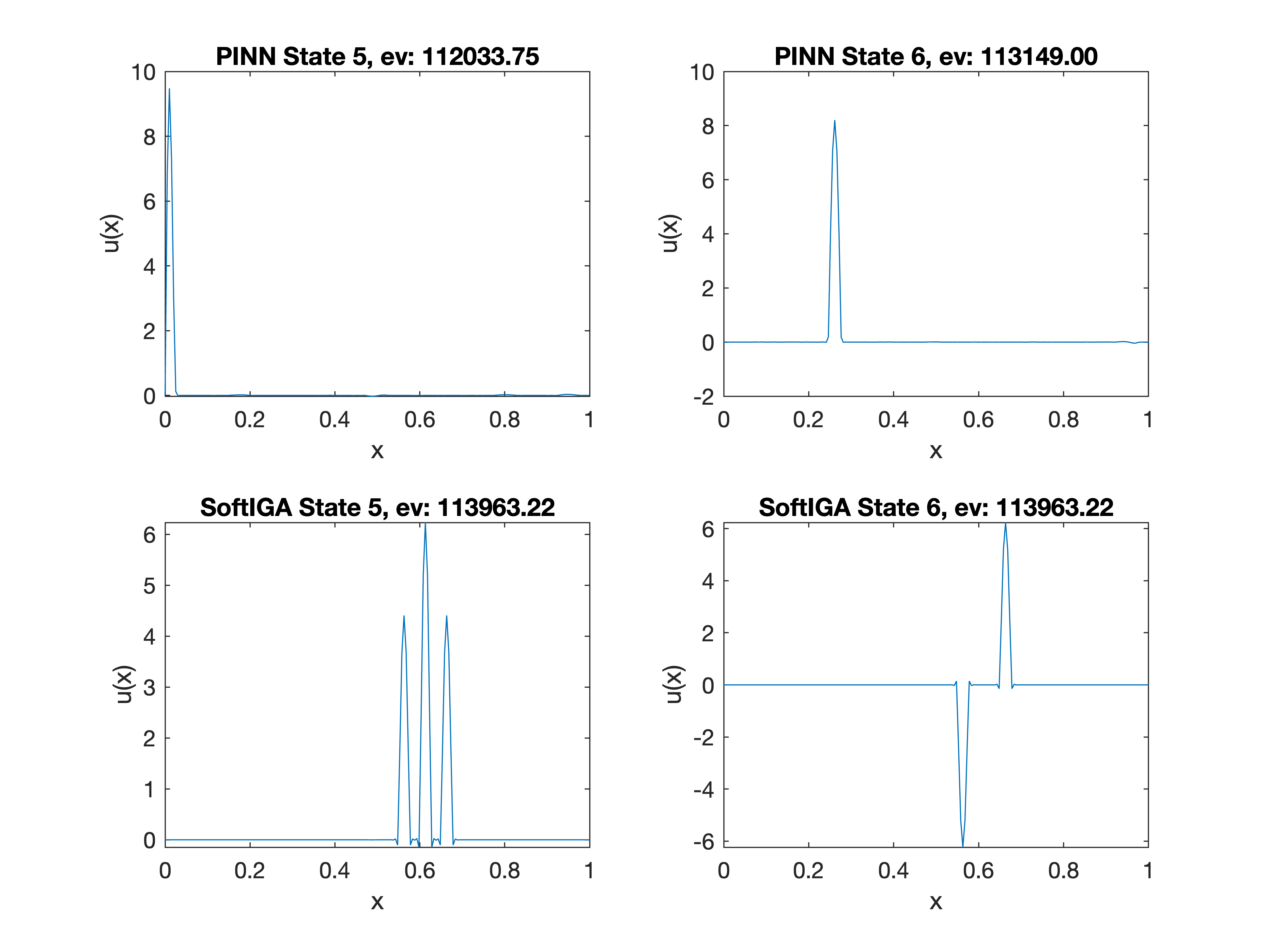}
	\caption{The PINN and SoftIGA predicted sixth and seventh eigenstates for the potential in \ref{fig:iga_pinn_V} with their respective eigenvalue calculation (ev).}
	\label{fig:iga_pinn_evecs}
\end{figure}

\rev{
Figure \ref{fig:iga_pinn_evecs1to4} plots the PINN's eigenstate prediction compared to the SoftIGA calculation for the case with a potential plotted in \ref{fig:iga_pinn_V}. We can see the PINN's eigenfunction matching well the ones calculated by SoftIGA, and the eigenvalues each with percentage errors less than 0.5\% from the SoftIGA eigenvalue (which is the same for all four of these eigenstates). However, Figure \ref{fig:iga_pinn_evecs} shows a significant difference between the PINN and SoftIGA calculation of the fifth and sixth eigenstates for this potential. The PINN's eigenvalue predictions are within 2\% of the SoftIGA calculations, however, the SoftIGA eigenfunctions are not localised to single regions. Instead, the eigenfunctions appear to be linear combinations of localised states, while the PINN's calculations are individual localised states. We believe that the individual localised states that make up the linear combination that SoftIGA finds, and the localised states that the PINN finds are all within the same eigenspace (have the same eigenvalue), and thus their linear combinations are also eigenstates with this eigenvalue. We believe this eigenspace is the span of all the localised states in the troughs of the potential that are one segment wide (for example, the intervals $[0, 0.02]$, $[0.08, 0.1]$, $[0.26, 0.28]$, $[0.38, 0.4]$, etc., in Figure \ref{fig:iga_pinn_V}). We also believe that the eigenstates in Figure \ref{fig:iga_pinn_evecs1to4} are all part of the eigenspace spanned by the localised states in the troughs of the potential that are two segments wide ($[0.16, 0.2]$, $[0.48, 0.52]$, $[0.78, 0.82]$, $[0.94, 0.98]$). Since our PINN works by scanning through each segment to specifically detect a localised state, the network does not pick up linear combinations of separate localised states, such as the SoftIGA calculations in Figure \ref{fig:iga_pinn_evecs}. Thus, our network is capable of separating these eigenspaces into their localised basis elements, while SoftIGA is not, displaying an area where PINNs can outperform traditional computational methods.
}

\color{black}

\subsection{An Example in Two Dimensions} \label{sec:2d}

In 2D, we consider a special case where the potential can be decomposed as $V(x,y) = V_1(x) + V_2(y)$ in a rectangular domain $\Omega = [0, 1]^2$. 
The Schrödinger operator can be rewritten as 
$$
- \Delta + V(x, y) = ( -\partial_{xx} + V_1(x) ) + (-\partial_{yy} + V_2(y) ),
$$
which allows the decomposition in the weak solution, leading to an equivalence of solving two 1D eigenvalue problems. 
In particular, in the finite and isogeometric element setting with tensor-product meshes, the 2D matrix eigenvalue problem can be also decomposed into tensor-product structures, leading to 1D solvers. 
In this setting, a base function in 2D can be rewritten as a product, i.e., $\phi(x,y) = \phi_1(x) \phi_2(y)$.
Following the derivations in \cite{calo2019dispersion,behnoudfar2020variationally}, 
for the eigenvalue problem \eqref{eq:vf} in 2D, let $W_h = \text{span}\{ \phi_{1,j} (x) \phi_{2,k} (y) \}$. 
Then, the bilinear forms can be decomposed as 

\begin{equation} \label{eq:sepb2d}
\begin{aligned}
a(\phi_{1,k}(x) \phi_{2,k}(y), \phi_{1,j}(x) \phi_{2,j}(y))  & = a_x(\phi_{1,k}(x), \phi_{1,j}(x) ) \cdot b_y(\phi_{2,k}(y),  \phi_{2,j}(y)) \\ 
& \quad + b_x(\phi_{1,k}(x), \phi_{1,j}(x) ) \cdot a_y(\phi_{2,k}(y),  \phi_{2,j}(y)), \\
b(\phi_{1,k}(x) \phi_{2,k}(y), \phi_{1,j}(x) \phi_{2,j}(y)) & = b_x(\phi_{1,k}(x), \phi_{1,j}(x) ) \cdot b_y(\phi_{2,k}(y),  \phi_{2,j}(y)),
\end{aligned}
\end{equation}
where the 1D bilinear forms are defined as 
\begin{equation}
\begin{aligned}
a_x(\phi_{1,k}(x), \phi_{1,j}(x) ) & = \int_0^1 \Big( \frac{\text{d}}{\text{d} x} \phi_{1,k}(x) \frac{\text{d}}{\text{d} x} \phi_{1,j}(x) + V_1(x) \phi_{1,k}(x) \phi_{1,j}(x)  \Big) \ \text{d} x ,  \\
\quad a_y(\phi_{2,k}(y),  \phi_{2,j}(y)) & = \int_0^1 \Big( \frac{\text{d}}{\text{d} y} \phi_{2,k}(y) \frac{\text{d}}{\text{d} y}  \phi_{2,j}(y) + V_2(x) \phi_{2,k}(y) \phi_{2,j}(y)  \Big) \ \text{d} y,\\
b_x(\phi_{1,k}(x), \phi_{1,j}(x) ) & = \int_0^1 \phi_{1,k}(x) \phi_{1,j}(x) \ \text{d} x, \\
b_y(\phi_{2,k}(y),  \phi_{2,j}(y)) & = \int_0^1 \phi_{2,k}(y)  \phi_{2,j}(y) \ \text{d} y.
\end{aligned}
\end{equation}

\begin{figure}[ht]
	\centering
	\includegraphics[width=.7\linewidth]{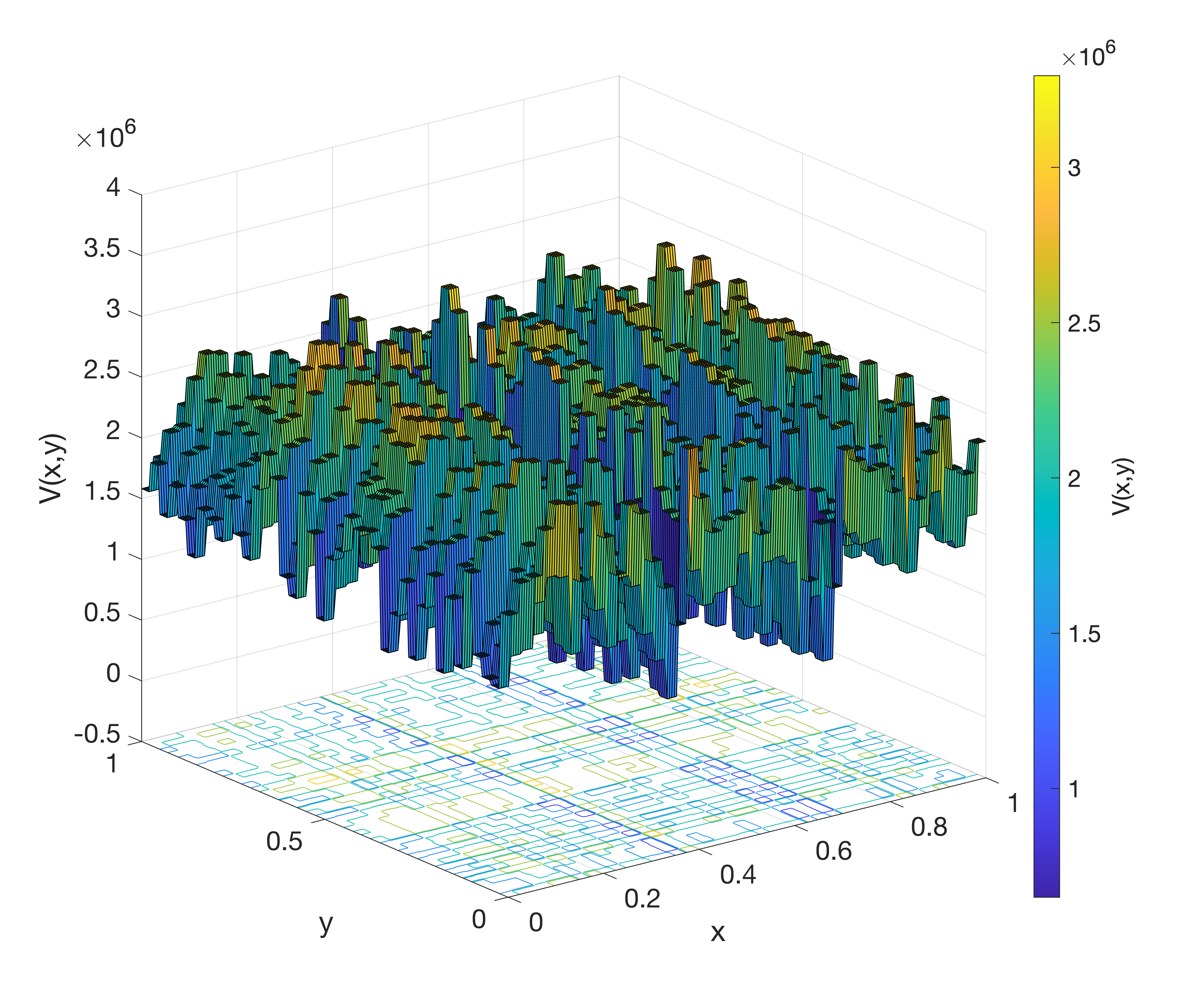}
	\caption{The surface plot of the potential $V(x,y) = V_1(x) + V_2(y)$ where $V_1$ and $V_2$ are separately randomly generated across $m=40$ regions using $n=200$ nodes according to the normal distribution with mean 1 and standard deviation 0.3, then scaled by $10^6$.}
	\label{fig:V_2D}
\end{figure}

\begin{figure}[ht]
	\centering
	\includegraphics[width=\linewidth]{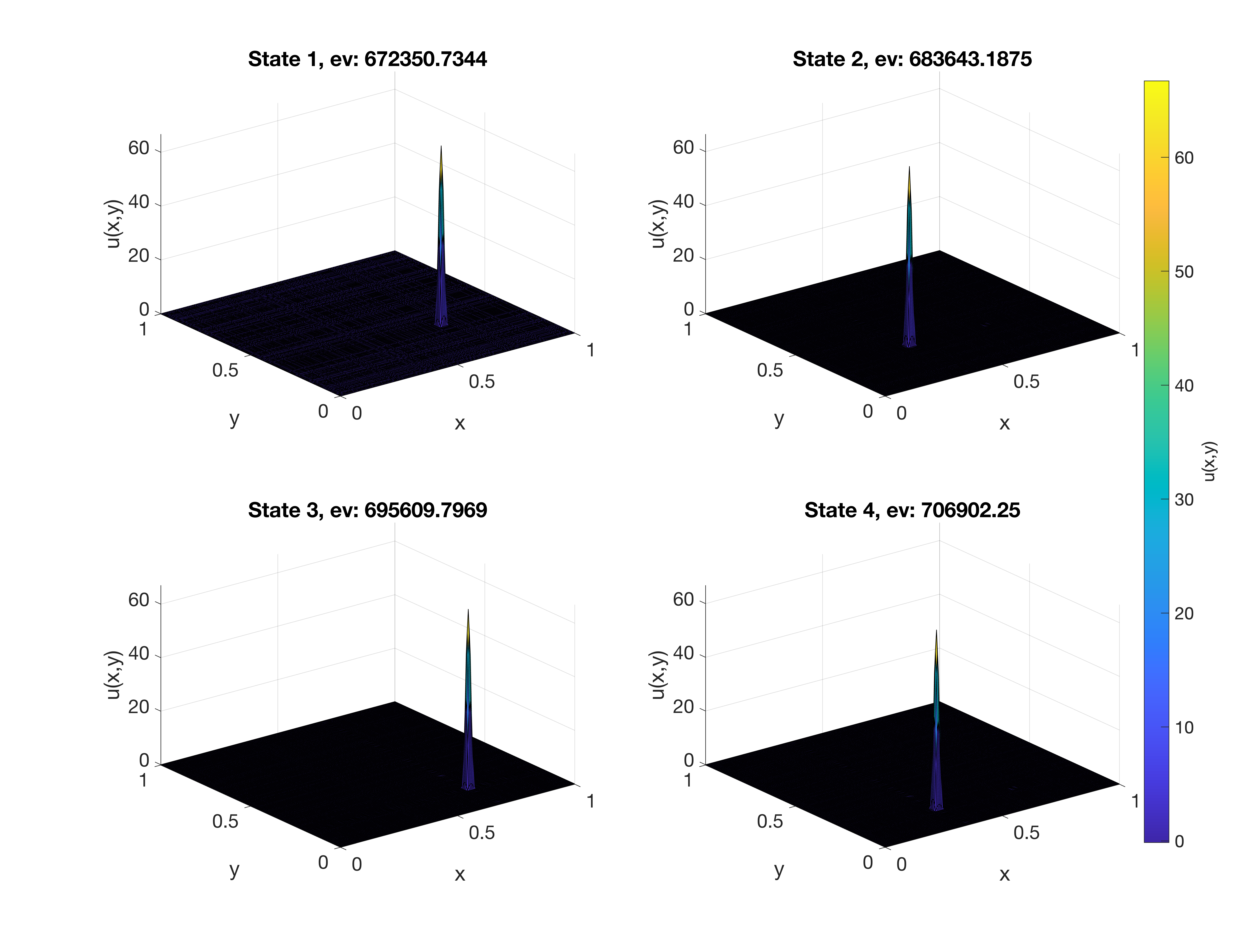}
	\caption{The first four PINN predicted eigenstates with the potential defined in Figure \ref{fig:V_2D}.}
	\label{fig:evecs_2D}
\end{figure}

This is also referred to as variational separability. With this in mind, we can rewrite the matrix eigenvalue problem \eqref{eq:mevp} in 2D as a problem with 1D tensor-products:
\begin{equation} \label{eq:sepmp2d}
(K_x \otimes M_y + M_x \otimes K_y) U = E_h (M_x \otimes M_y) U,
\end{equation}
which allows 
\begin{equation}
\big( (K_x  - E_x M_x) \otimes M_y + M_x \otimes (K_y - E_y M_y) \big) U = 0,
\end{equation}
where $E_h = E_{h,x} + E_{h,y}$. As the dimensions are separated, this leads to two 1D eigenvalue problems (see \cite{ainsworth2010optimally,calo2019dispersion} for more details).
\begin{equation} \label{eq:2d-1d}
\begin{aligned}
K_x U_x & = E_x M_x U_x, \\
K_y U_y & = E_y M_y U_y. \\
\end{aligned}
\end{equation}
With this insight of dimension decomposition in mind, we apply the PINN in $x$ and $y$ dimensions to find eigenstates for the 2D problem. 
The eigenvalues in 2D are the sum of the 1D eigenvalues while the eigenstates in 2D are the tensor-products of the 1D eigenstates.
This significantly reduces the computational costs.
As an example, Figure \ref{fig:V_2D} shows the surface plot of one of these two-dimensional potentials, with the PINN eigenstates displayed in Figure \ref{fig:evecs_2D}.
When the two 1D potentials admit localised eigenstates, the eigenstates of the constructed 2D potential are also localised, as $u(x,y) = u_1(x)u_2(y)$ is only nonzero on the small rectangular region where both $u_1(x)$ and $u_2(y)$ are nonzero. This is demonstrated by Figure \ref{fig:evecs_2D}, showing the eigenstates localised to the rectangular regions.

\color{black}

%
%
%
%
%

\section{Concluding Remarks} \label{sec:conc}

Recently, there has been a growing interest in using machine learning and neural networks to solve differential equations, particularly in the study of Hamiltonians. In this paper, we extend current models by constructing a neural network capable of distinguishing eigenstates from Hamiltonians with randomly distributed potentials that lead to localisation in the states. Our approach involves incorporating a normalisation loss term to avoid trivial solutions and an orthogonality loss term to search for higher eigenstates. Our method eliminates the need for driving terms, which can hinder a network's ability to converge to an admissible solution. The novelty of our approach lies in a new localisation loss term, which encourages the network to converge to a solution that is localised to a specific region. This idea is based on the physical fact that localised eigenstates occur in randomly distributed potentials. We utilise the network's initial attempt at convergence to identify regions of localised states before adding in this term to achieve the desired approximation. We demonstrate the effectiveness of our approach in discovering eigenstates for potentials generated randomly according to different distributions, with the highlight being the Bernoulli distribution, which leads to eigenstates with identical eigenvalue. Our model successfully predicts these eigenstates with good accuracy, overcoming a challenge faced by current models.

There are several potential avenues for future work. Some potential future directions in this research area include:
(1) exploring the use of more complex loss functions that can capture additional physical constraints, such as symmetries or conservation laws;
(2) investigating the use of different network architectures, such as convolutional neural networks or attention-based models, to improve accuracy and scalability;
(3) generalising the approach to higher-dimensional systems or non-linear differential equations;
and (4) investigating the performance of the approach with larger and more complex systems, including systems with a larger number of particles or more complex interactions.
Overall, the use of ANNs to solve differential equations and approximate eigenstates of Hamiltonians has shown to be a promising approach, and there are many potential directions for future research in this area.


\bibliography{ref}


\end{document}